\documentclass[12pt,preprint]{aastex}
\usepackage{emulateapj5}
\usepackage{epsfig}

\def\vkm{km s$^{-1}$}
\def\vkme{\textrm{km s}^{-1}}
\def\degree{$^\circ$}
\def\arcs#1{$#1''$}
\def\arcsa#1#2{$#1^{\prime\prime}_{^\textrm{.}}#2$}
\def\arcsaq#1#2{#1^{\prime\prime}_{^\textrm{.}}#2}
\def\solarmass{$M_\odot$}

\def\Jyb{Jy beam$^{-1}$}
\def\mJyb{mJy beam$^{-1}$}
\def\Jybk{Jy beam$^{-1}$ km s$^{-1}$}
\def\tlabel#1{(\textit{#1})}

\def\cmc{cm$^{-3}$}
\def\cms{cm$^{-2}$}
\def\micron{$\mu$m}
                                                                                
\def\ra#1#2#3#4{#1^\mathrm{h} #2^\mathrm{m} #3^\mathrm{s}_{^\textrm{.}} #4}
\def\dec#1#2#3#4{#1\degr #2\arcmin #3^{\prime\prime}_{^\textrm{.}}#4}

\def\H2{H$_2$}
\def\N2HP{N$_2$H$^+$}
\def\HCOP{HCO$^+$}
\def\cCO{C$^{18}$O}

\def\bCO{$^{13}$CO}
\def\NH3{NH$_3$}

\def\SOt{$N_J=8_9-7_8$}
\def\SOta{$N_J=5_6-4_5$}
\def\HCOP{HCO$^+$}
\def\aHCOP{H$^{13}$CO$^+$}
                                                                                
\def\putfig#1#2#3{\epsfig{scale=#1,angle=#2,figure=#3}}
\def\leftblank#1{}

\begin{document}

\title{HH 212: SMA observations of a remarkable protostellar jet}
\author{Chin-Fei Lee\altaffilmark{1}, 
Paul T.P. Ho\altaffilmark{2,3}, Henrik Beuther\altaffilmark{4}, 
Tyler L. Bourke\altaffilmark{3}, Naomi Hirano\altaffilmark{2}, 
Hsien Shang\altaffilmark{2}, and Qizhou Zhang\altaffilmark{3}
}
\altaffiltext{1}{
Harvard-Smithsonian Center for Astrophysics, Submillimeter Array,
645 North A'ohoku, Hilo, HI 96720; cflee@cfa.harvard.edu}
\altaffiltext{2}{Academia Sinica Institute of Astronomy and Astrophysics,
P.O. Box 23-141, Taipei 106, Taiwan}
\altaffiltext{3}{Harvard-Smithsonian Center for Astrophysics, 60 Garden
Street, Cambridge, MA 02138}
\altaffiltext{4}{Max-Planck-Institute for Astronomy, K\"onigstuhl 17,
69117 Heidelberg, Germany}

\begin{abstract}
HH 212 is a nearby (460 pc) protostellar jet discovered in \H2{} 
powered by a Class 0 source, IRAS 05413-0104, in the L1630 cloud of Orion.
It is highly collimated and symmetric 
with matched pairs of bow shocks on either side of the source.
We have mapped it in 850 \micron{} continuum, SiO ($J=8-7$), CO ($J=3-2$), 
SO (\SOt{}), \HCOP{} ($J = 4-3$), and \aHCOP{} ($J = 4-3$)
emission simultaneously at $\sim$ \arcs{1} resolution with the Submillimeter 
Array (SMA). 
Thermal dust emission is seen in continuum around the source, 
mainly arising from an inner envelope (i.e., the inner
part of a previously seen flattened envelope) and a possible disk.
The inner envelope is also seen with rotation in CO, 
\HCOP{}, and probably SO.
Like \H2{} emission, CO and SiO emission are seen
along the jet axis but extending closer to the source,
tracing the bow shocks with a broad range of velocities and
the continuous structures in between.
SO emission is seen only around
the source, forming a jetlike structure extending along the jet axis 
from the source, likely tracing the jet near the launching region.
The jet is episodic and bending. It may also be slightly precessing as
the jetlike SO structure shows a slight
S-shaped symmetry about the source. A hint of jet rotation is also
seen across the jet axis. 
Internal outflow shells are seen in CO and
\HCOP{}, associated with the bow shocks in the inner part of the
jet. The bases of the \HCOP{} shells are seen with a hint of rotation
similar to that seen in the inner envelope, probably consisted mainly of the
material extended from the inner envelope and even the possible disk. 
The bases of the outflow shells are also seen
in \aHCOP{} and even the continuum, probably tracing the dense
material extended from around the same regions.
Outflow shells are also seen in CO surrounding the jet extending
out from the \H2{} nebulae seen around the source.
\end{abstract}


\section{Introduction}

Protostellar jets are seen associated with
low-mass protostars in the early stages of star formation.
In spite of numerous studies,
their physical properties (e.g., speed, episodic nature, 
collimation, and angular momentum)
are still not well understood \citep{Hartigan2000}.
They are believed to be launched from accretion disks around the protostars
\citep{Shu2000,Konigl2000},
allowing us to probe the accretion process, 
which remains heretofore unresolved, as it requires us to observe
directly in the inner parts at the AU scale.
The Submillimeter Array (SMA)\footnote{ The Submillimeter Array is a joint project between the
Smithsonian Astrophysical Observatory and the Academia Sinica Institute of
Astronomy and Astrophysics, and is funded by the Smithsonian Institution and
the Academia Sinica.} \citep{Ho2004}, 
with the capability to probe warm and dense molecular gas at high angular
resolution, can be and has been used to
study the physical properties of the jets in great detail
\citep{Hirano2006,Palau2006}.

The HH 212 jet, which is nearby (460 pc) in the L1630 cloud of Orion
and remarkably well-defined,
is one of the best 
candidates to investigate the physical properties of the jets.
It was discovered in shock excited \H2{} emission
\citep{Zinnecker1998},
powered by a low-luminosity ($\sim$ 14 $L_\odot$) 
Class 0 source (or protostar), IRAS 05413-0104. It is highly collimated
and highly symmetric with matched pairs of bow shocks on either side of
the source.
Water masers are seen at the bases of the jet with proper motions
along the jet axis \citep{Claussen1998}.
The jet interacts with the ambient material, driving a 
collimated CO outflow around it 
(Lee et al. 2000; Lee et al. 2006, hereafter \citet{Lee2006}).
Deep observations in \H2{} using the Infrared Spectrometer And Array Camera
(ISAAC) of the ESO Very Large Telescope (VLT)
show a pair of diffuse nebulae near the bases of the jet,
probably tracing the outflow cavity walls illuminated by the bright
bow shocks around the source \cite[][hereafter MZAML02]{McCaughrean2002}.
The inner part of the jet is also seen in
shock excited SiO emission \citep{Chapman2002,Gibb2004,Takami2006} 
and high-velocity CO emission \citep{Lee2006}.
Since the jet is compact, dense, and warm, high-resolution observations
in high-$J$ transition lines of SiO, CO, and other shock tracers
are presented here to study its physical properties.

\section{Observations}\label{sec:obs}

Observations toward the HH 212 jet were carried out with the SMA
on December 2, 2005 in the compact
configuration and on January 14, 2006 in the extended configuration.  SiO
($J=8-7$), CO ($J=3-2$), SO (\SOt{}), \HCOP{} ($J = 4-3$), and \aHCOP{} 
($J = 4-3$) lines were observed simultaneously with 850 \micron{} continuum using the
345 GHz band receivers. 
The rest frequency, upper energy level, and critical density
of these lines are given in Table \ref{tab:obs}.
The receivers have two sidebands, lower and upper,
covering the frequency range from 345.5 to 347.5 and from 355.5 to 357.5 GHz,
respectively. Combining the two sidebands results in
a total bandwidth of 4 GHz centered
at 351.5 GHz (or $\lambda \sim$ 850 \micron{}).
Seven antennas were used in the array, giving baselines with projected
lengths ranging from 8.5 to 180 m.
Since the primary beam has a size of $\sim$ \arcs{35}, two pointings
with a separation of \arcs{18} were used to map
the inner part of the jet within \arcs{25} from the source.
For the correlator,
256 spectral channels were used for each 104 MHz chunk,
resulting in a velocity resolution of $\sim$ 0.35 \vkm{} per channel.

The visibility data were calibrated with the MIR package,
with Saturn, Uranus, and Quasar 3C454.3 as passband calibrators, 
Quasar J0530+135 as a gain calibrator, and Uranus as a flux calibrator.
The calibrated visibility data were imaged with the MIRIAD package.
The dirty maps that were produced from the calibrated visibility data
were CLEANed using the Steer clean method,
producing the CLEAN component maps.
The final maps were obtained by restoring the CLEAN component
maps with a synthesized 
(Gaussian) beam fitted to the main lobe of the dirty beam. 
With natural weighting, the synthesized beam has 
a size of \arcsa{1}{16}$\times$\arcsa{0}{84} 
at a position angle (P.A.) of 38\degree{}.
In some cases, however, in order to study the compact
structures (i.e., the jet and the possible disk) in the system,
the visibilities with $uv$ radius shorter than 40 m
(or $\gtrsim$ \arcsa{2}{5}) are excluded, resulting in a 
smaller synthesized beam with a size of \arcsa{0}{96}$\times$\arcsa{0}{69}
at P.A.$=$ 38\degree{}.
The rms noise level is $\sim$ 0.16 \Jyb{} in the channel maps
and 5 \mJyb{} in the continuum map.
The velocities of the channel maps are LSR.
The absolute positional accuracy in our maps is expected to be
one tenth of the synthesized beam, or $\sim$ \arcsa{0}{1}.

\section{Results}

In the following, our results are presented with the \H2{} image
made with the VLT (MZAML02)
%with data from October 2000, October 2001, and January 2002
for comparison.
The \H2{} image, which shows a clear jet consisting of a chain 
of bow shocks and sinuous continuous structures in
between in great detail,
is used as a reference for the jet.
Since our observations were carried out about 4 years later than
the \H2{} image, the bow shocks, which
were found to be moving at 100$-$200 \vkm{} or 
\arcsa{0}{045}$-$\arcsa{0}{09} per year (MZAML02),
might have moved down along the jet axis by \arcsa{0}{18}$-$\arcsa{0}{36}
from their positions in that image.  This movement of the bow shocks,
however, does not affect significantly our comparison and conclusions,
considering that the angular resolution of our observations is $\sim$
1\arcsec{} along the jet axis.
The \H2{} jet is almost in the plane of the sky, with the 
blueshifted side to the north and the redshifted
side to the south of the source \citep{Zinnecker1998}.
The systemic velocity in this region is assumed to be
$1.7\pm0.1$ \vkm{}, as in \citet{Lee2006}.
Throughout this paper, the velocity is relative to this systemic value.

\subsection{850 \micron{} Continuum Emission} \label{sec:cont}

\begin{figure} [!hbp]
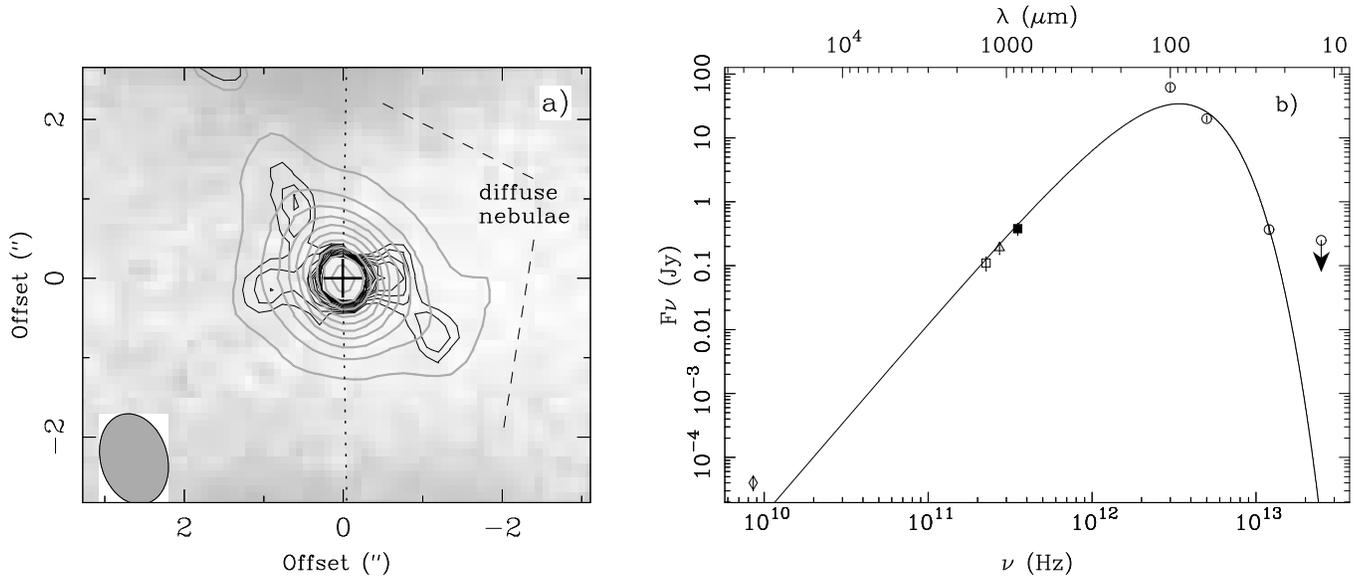

\centering
\putfig{0.75}{270}{f1.ps}
\figcaption[]
{
\tlabel{a} 850 \micron{} continuum contours plotted on top of the 
\H2{} image adopted from MZAML02. Note that the image has
been rotated by 22.5\degree{} clockwise.
Gray contours are from the restored map with a synthesized beam of
\arcsa{1}{16}$\times$\arcsa{0}{84}, going from 4 to 40 
$\sigma$ with a step of 4 $\sigma$, where $\sigma = 5$ \mJyb{}. 
Black contours are from the CLEAN component map.
The cross marks the source position.
\tlabel{b} A single-temperature fit to the spectral energy distribution
of the continuum source (see text for detail). 
The squares are from our SMA observations, with the open one
from \citet{Lee2006} and the filled one from this paper.
The open circles, triangle, and diamond are from the IRAS, JCMT 
\citep{Zinnecker1992}, and VLA \citep{Galvan2004} observations, respectively.
\label{fig:cont}}
\end{figure}

A continuum source is detected at 850 \micron{} 
in between the diffuse \H2{} nebulae
with an emission peak at
$\alpha_{(2000)}=\ra{05}{43}{51}{404}\pm\arcsaq{0}{1}$,
$\delta_{(2000)}=\dec{-01}{02}{53}{10}\pm\arcsaq{0}{1}$, 
and a total integrated flux of 0.38$\pm0.10$ Jy (Fig. \ref{fig:cont}a).
This peak position is within \arcsa{0}{1} from that
found at $\lambda= 3.5$ cm with the VLA at an
angular resolution of \arcsa{0}{3} \citep{Galvan2004}, and is thus
considered as the source position throughout this paper.
The structure of the continuum source is not well resolved in the restored 
map.
It is better seen in the CLEAN component map, which
shows a faint flattened structure 
with a radius of $\sim$ \arcs{1} (460 AU) 
perpendicular to the jet axis and a bright unresolved compact structure
with a radius $<$ \arcsa{0}{3} (138 AU)
at the center. 
Faint emission is also seen extending to the northeast and southwest.
The faint flattened structure, 
which is not resolved in the minor
axis, may trace the inner envelope, i.e., the inner part of the
edge-on flattened envelope seen in \citet{Lee2006}, while
the compact structure may trace the envelope further in and
a possible disk. 
The faint emission extending to the northeast and southwest, 
on the other hand, may trace the material extended from the
inner envelope,
as discussed later in \S \ref{sec:shells}.
A compact source with an envelope is also seen in the 
amplitude versus $uv$ distance plot.
However, observations at higher angular resolution are really needed to
confirm the structure seen in the CLEAN
component map.
The spectral energy distribution (SED) of the continuum source
\cite[see Fig. \ref{fig:cont}b and also][]{Lee2006}
indicates that the continuum emission at 850 \micron{} is 
mainly thermal dust emission. 
Assuming a constant temperature, a mass opacity
$\kappa_\nu = 0.1 (\nu/10^{12} \textrm{\scriptsize Hz})^\beta$
cm$^2$ g$^{-1}$ \citep{Beckwith1990},
and a source size of 1 arcsec$^2$ for the dust,
the SED can be fitted with $\beta=1$ and 
a temperature of $\sim$ 48 K.
Thus, the (gas $+$ dust) mass is estimated to be $\sim$ 0.08 $M_\odot$,
as in \citet{Lee2006}.
With $\beta=1$, the bright compact structure at the center may indeed
harbor a disk around the source \citep{Jorgensen2006}.

\subsection{Jet Axis and Bending}\label{sec:jetaxis}

The northern component and southern component of the \H2{} jet are not
exactly antiparallel (Figs. \ref{fig:jet}a and \ref{fig:jet}b). Their inner
parts are misaligned by $\sim$ 2\degree{}, with their axes found to have a
P.A. of 21.5\degree{}$\pm0.5$\degree{} and
203.5\degree{}$\pm0.5$\degree{}, respectively, by connecting the source to
the inner \H2{} bow shocks out to bow shocks NK7 and SK7. 
Since their
original paths of motion are likely to be antiparallel, this misalignment
suggests a presence of a jet bending (see \S \ref{sec:dis_bend} for possible
mechanisms).
Assuming both bent by the same degree,
their inner parts are both bent by $\sim$ 1\degree{} to the west
%at a P.A. of $-$67.5\degree{}$\pm1$\degree{}, 
with the original axis having a P.A. of 22.5\degree{}$\pm1$\degree{}.
Note that the bending actually continues further out and beyond
bow shocks NB1/2 and SB1/2.

The outer parts of the jet
may have different original axis and bending because the
tips of bow shocks NB3 and SB4 are on the opposite sides of the axes of the
inner parts. The original axis, which can be estimated by connecting the
tips of bow shocks NB3 and SB4, is found to have a P.A. greater than that of
the inner parts, suggesting that the jet axis might have rotated clockwise
more than a few thousands years ago. However, without knowing
the source position at the time when bow shocks NB3 and SB4 were formed, we
are not able to determine the bending.

\subsection{SiO, CO, and SO emission}

\subsubsection{Morphologies}

SiO emission is detected mainly along the jet axis (Fig. \ref{fig:jet}c). 
About 70\% of the single-dish flux is recovered from our observations,
comparing to the Atacama Submillimeter Telescope Experiment (ASTE) 
observations \citep{Takami2006}.
It is detected toward the bow shocks and the continuous structures
seen in \H2{}.
Bow shocks NK2, NK4, SK2, and SK4, which are
bright in \H2{}, are also bright in SiO.
However, bow shocks NK1 and SK1, which are bright and prominent in \H2{},
are faint in SiO and only detected on their eastern sides.
SiO emission is also
detected at $\sim$ \arcs{1} north and south of the source, forming knotty
structures labeled knots SN and SS, respectively (see Fig. \ref{fig:injet}a
for the closeup). The knots are spatially unresolved
with a diameter (transverse size) of less than \arcs{1} (460 AU). 
A faint elongated structure is seen extending to the southwest 
from knot SS, fitting in between the wings of bow shock SK1, 
possibly tracing the ``intrinsic'' jet itself, but not the shock.

\begin{figure} [!hbp]
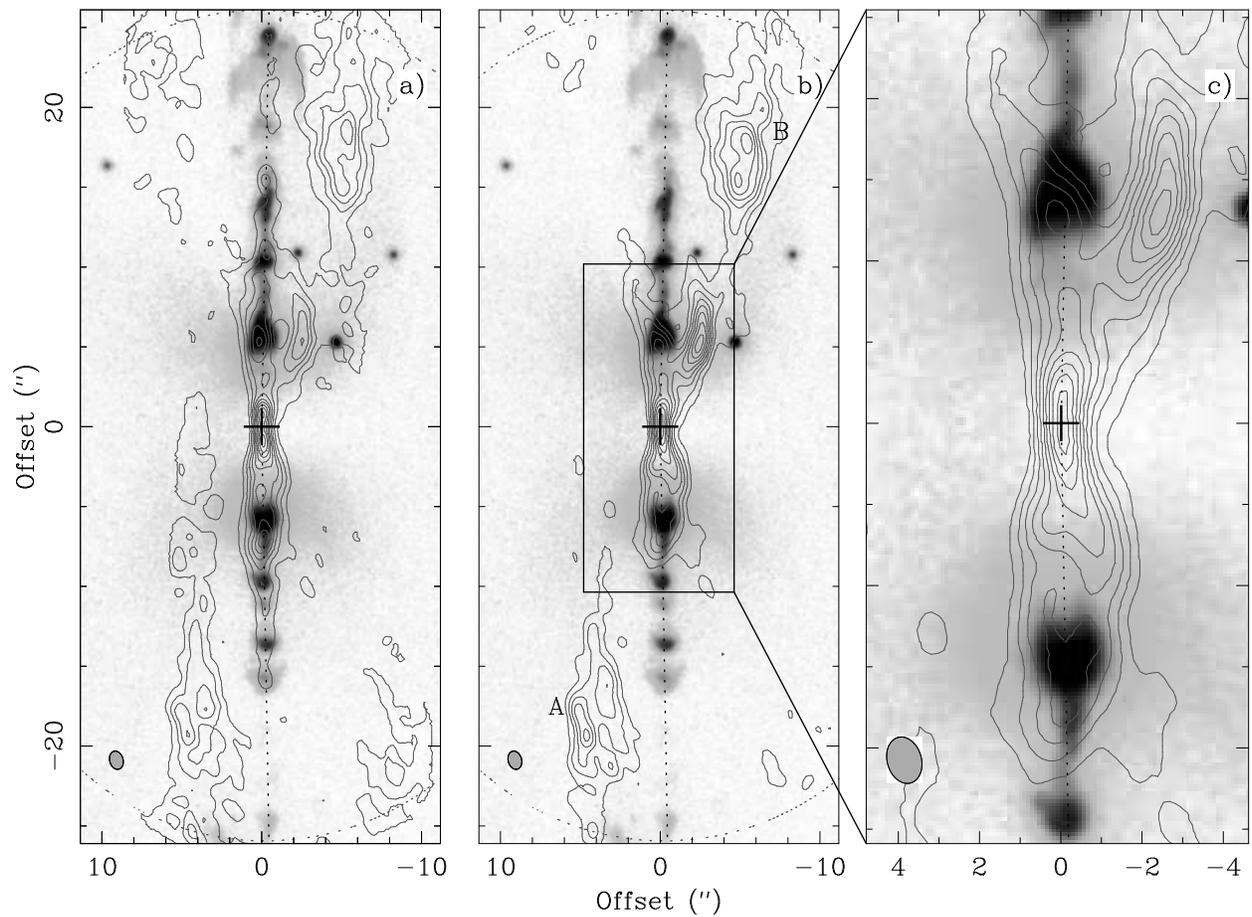

\centering
\putfig{0.7}{270}{f4.ps}
\figcaption[]
{CO contours plotted on top of the \H2{} image.
The synthesized beams all have a size
of \arcsa{1}{16}$\times$\arcsa{0}{84}.
\tlabel{a} shows the total CO emission integrated from -18.4 to 13.3 \vkm{}.
The contours go from 4 to 36
$\sigma$ with a step of 4 $\sigma$, where $\sigma = 0.55$ \Jybk{}.
\tlabel{b} and \tlabel{c} show the low-velocity CO emission integrated
from  $-$4.0 to 3.1 \vkm{}.
The contours go from 5 to 45 $\sigma$ with a
step of 5 $\sigma$, where $\sigma = 0.24$ \Jybk{}.
\label{fig:COshell}
}
\end{figure}

CO emission is detected not only
along the jet axis but also toward outflow shells (Fig.
\ref{fig:COshell}a).
In the following, two velocity ranges, high and low, are selected to
show these two components separately.
At high velocity (from $-$18.4 to $-$7.1 \vkm{} and from
3.4 to 13.3 \vkm{}),
CO emission is seen mainly along the jet axis
toward the bow shocks and the continuous structures seen in \H2{},
with some coincident with the SiO emission 
(compare Figs. \ref{fig:jet}c and \ref{fig:jet}d).
CO emission is also detected at $\sim$ \arcs{1} away from the source (see
Fig. \ref{fig:injet}b for the closeup), 
associated with knots SN and SS seen in SiO.
At low velocity (from $-$4.0 to 3.1 \vkm{}), 
limb-brightened outflow shells (labeled A and B) are seen
extending to the north and south, respectively, from the \H2{} 
nebulae (Fig. \ref{fig:COshell}b).  They seem to connect to
the \H2{} bow shocks SB1/2, NK7, and NB1/2 and some further down the jet
axis, and are thus probably internal outflow shells driven by them.
Throughout this paper, outflow shells are called internal if not driven by the
leading bow shocks at the heads of the jet.
%Like those \H2{} bow shocks, the
%shells are bright on the west in the north and on the east in the south.
Near the source, a
limb-brightened internal outflow shell is seen in the south
connecting to bow shock SK1 (Fig. \ref{fig:COshell}c).
Internal outflow shells are also seen in the north 
but associated with more than one bow shock.
The emission extending along the jet axis to bow shock NK1
is unresolved but may trace an internal outflow shell as a counterpart 
of that extending to bow shock SK1.

SO emission, on the other hand, is detected mainly around the source (Fig.
\ref{fig:injet}c).
Jetlike emission is seen along the jet axis extending from the source to the
north and south to knots SN and SS, respectively,
likely tracing the jet near the launching region.
The northern component is shifted
slightly to the east and the southern component is shifted slightly to the
west of the jet axis, showing a slight S-shaped symmetry about the source.
%Some of the SO emission may arise from a disk, with a peak at the source.

\subsubsection{Kinematics}

\begin{figure} [!hbp]
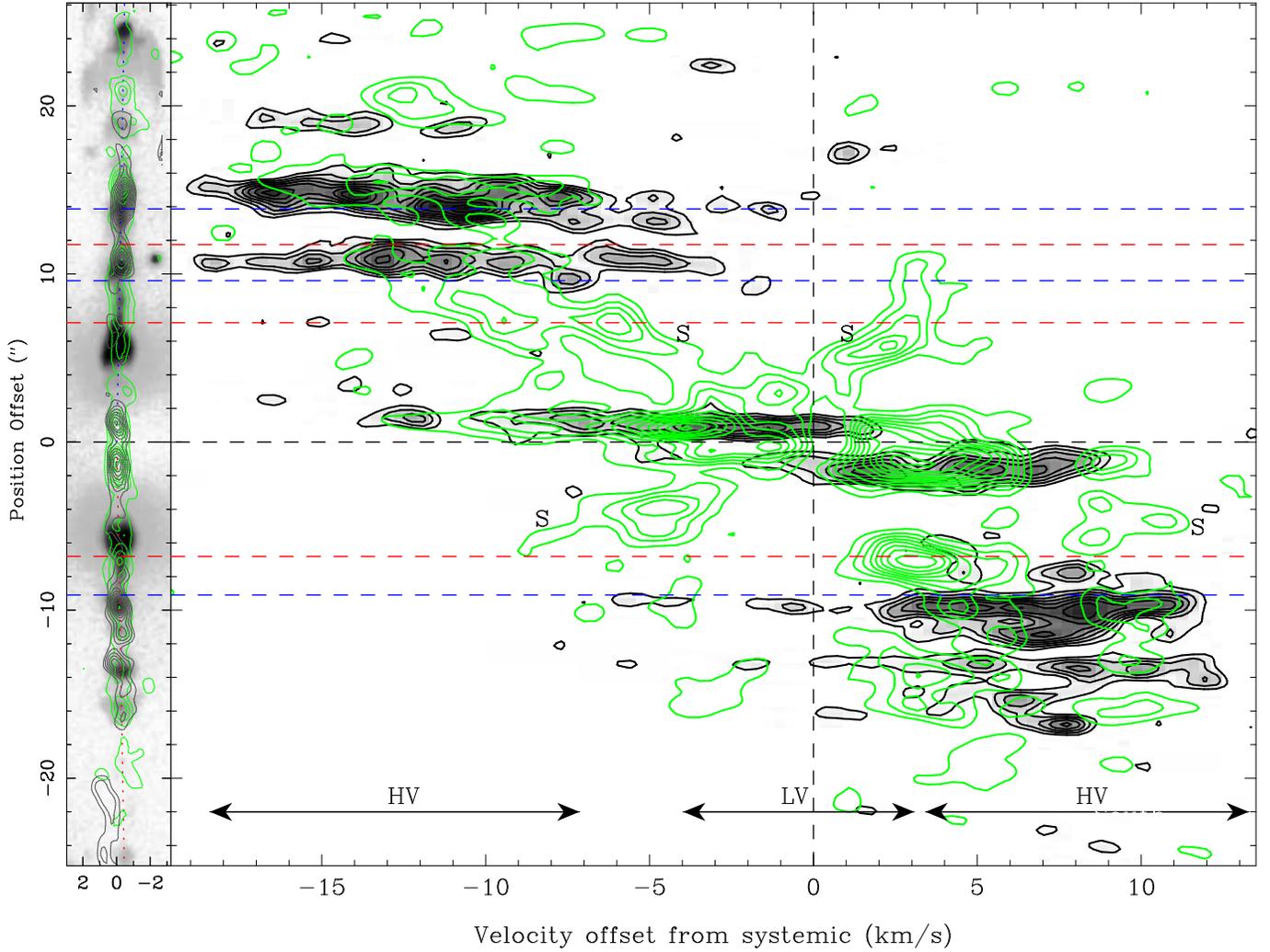

\centering
\putfig{0.7}{270}{f5.ps}
\figcaption[]
{PV diagrams of the SiO and CO emission cut along the jet axis
with a width of \arcsa{0}{9}.
\tlabel{a} SiO (gray) and high-velocity CO (green) contours on top of the \H2{} image.
\tlabel{b} PV diagrams of the SiO (black contours with image) and CO 
(green contours) emission.  Here, HV and LV denote the high-velocity and
low-velocity ranges, respectively, for the CO emission. 
The red and blue
dashed lines mark the near and far ends 
of the continuous structures.
\label{fig:pvSiOCO}
}
\end{figure}

The kinematics of the bow shocks and the continuous structures can be
studied with the position-velocity (PV) diagrams 
of the SiO and CO emission cut along the jet axis (Fig. \ref{fig:pvSiOCO}). 
In the PV diagrams, the CO emission around the systemic
velocity, which merges with that of the ambient cloud, and the CO emission from
7 to 9 \vkm{}, which merges with that of the foreground ambient cloud
with the same velocity range, are resolved out from our observations 
\cite[see also][]{Lee2006}.
The internal outflow shells (labeled S) near the source are seen in
CO with the velocity magnitude increasing with the distance from the source
toward bow shocks NK1, NK2, and SK1 on both the redshifted and blueshifted
sides, indicating that the shells around the bow shocks 
are expanding mainly transversely 
perpendicular to the jet axis.
This is expected for outflow shells 
driven by jet-driven bow shocks \cite[see, e.g.,][]{Lee2001}.
The bases of the shells, on the other hand, may also
have velocity component parallel to the jet axis.

Like the \H2{} emission, the SiO and CO emission toward the bow shocks and
the continuous structures are mainly blueshifted in the
north and redshifted in the south.
They are, however, more blueshifted in the north 
than redshifted in the south, with a mean (representative) velocity
of $\sim$ $-11$ \vkm{} in the north while 7 \vkm{} in the south,
similar to that seen in the \H2{} emission \citep{Takami2006}.
The reason for this asymmetry in the velocity is unclear.
The SiO emission is seen with a range of velocities toward the bow shocks:
bow shocks NK2, NK4, SK2, and SK4 are bright and 
seen with a velocity range of $\sim$ 15 \vkm{},
while other bow shocks are faint and seen with a narrower 
velocity range.
The SiO emission is not well resolved,
probably some from the bow tips with a broad range of
velocities and some from the bow wings with the velocity magnitude
decreasing rapidly away from the bow tips. 
%In addition, the SiO emission toward bow shock NK4 may arise from both of these
%components.
The CO emission is also seen with a range of velocities toward the bow
shocks, some with similar and some with narrower velocity range than that of
the SiO emission.
Some CO emission is probably from the bow tips and some from the bow wings,
with some (e.g., bow shocks NK2 and SK2) coincident with the SiO emission.
Note that
a broad range of velocities is also seen ahead of bow shock NK4 in CO,
suggesting that a shock is formed there.

The continuous structures between the bow shocks may trace
the ``intrinsic'' jet itself
and thus can be used to study the jet kinematics.
In the south, the continuous structure is seen
between bow shocks SK1 and SK2 in CO and SiO.
The SiO and CO emission there together may show that its velocity
increases with the distance from bow shocks SK1 to SK2.
In the north, the continuous structures are seen between bow shocks NK1 and
NK2 and between bow shocks NK2 and NK4 in CO. Their velocity also seems to
increase with the distance from one bow shock to the next.

\begin{figure}[!hbp]
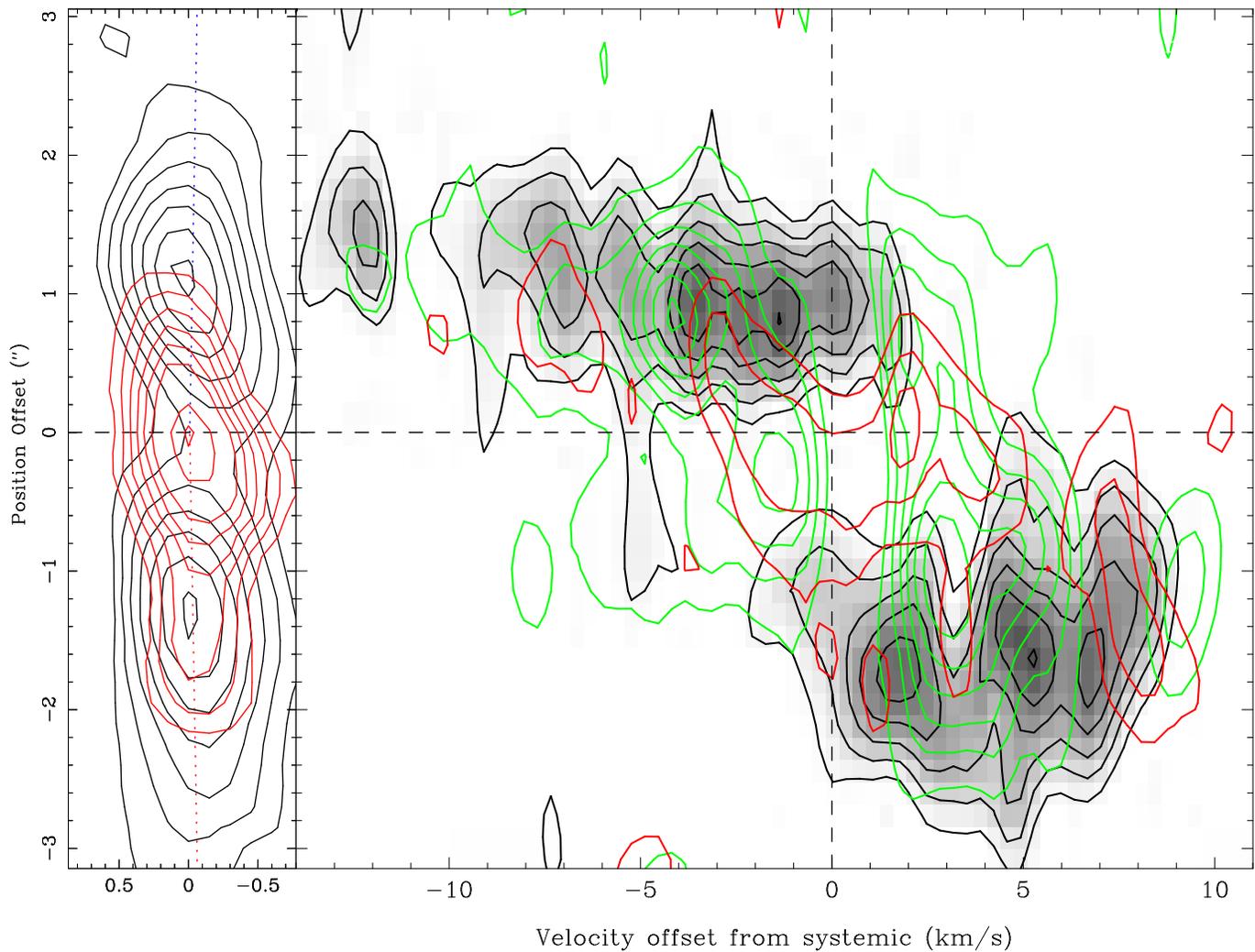

\centering
\putfig{0.7}{270}{f6.ps}
\figcaption[]{
PV diagrams of the SiO, CO, and SO emission cut along the jet axis
with a width of \arcsa{0}{9}.
\tlabel{a} SiO (black) and SO (red) contours.
\tlabel{b} PV diagrams of the SiO (black contours with image),
CO (green contours), and SO (red contours) emission. 
\label{fig:pvSiOCOSO}
}
\end{figure}

The kinematics of knots SS and SN around the source can be studied with
the PV diagrams of the SiO, CO, and SO emission 
(Fig. \ref{fig:pvSiOCOSO}). 
In SiO, knots SN and SS are also seen with a broad range of velocities, 
but with a lower mean velocity than that seen toward the bow shocks 
further away.
The kinematics of the CO emission is unclear, probably with a velocity magnitude increasing
with the distance from the source toward the knots on both blueshifted
and redshifted sides.
The kinematics of the SO emission is also unclear, probably partly similar to
that of the SiO emission and partly similar to that of the CO emission.
Observations at higher resolution are really needed to resolve their
structures and study the detailed relationship among them.

\subsection{\HCOP and \aHCOP emission}

\subsubsection{Morphologies}\label{sec:morhcop}

\HCOP{} emission is seen extending to the north
from the source surrounding the CO emission
that extends to bow shock NK1
(Fig. \ref{fig:hcop}a).
The emission is likely from a limb-brightened shell as seen in the blueshifted emission,
but with the east side brighter than the west side (Fig. \ref{fig:hcop}b).
It extends to the faint bow shock NF and then
to the wings of bow shock NK1.
Its base is also seen in continuum coincident with its
northeast extension (Fig. \ref{fig:hcop}c).
In the south, a similar shell is expected to be seen in \HCOP{}, but it is
faint and only seen at the base surrounding the CO shell that extends to bow
shock SK1. As in CO, the shell is brighter on the west side. 
Note that
the emission (labeled NW in Fig. \ref{fig:hcop}c) 
extending to the northwest from the source is
almost perpendicular to the jet axis coincident with the continuum emission
and thus may arise from the inner envelope.

A shell is also seen in \aHCOP{} extending to the north from the source,
surrounding the east side of the north base of the \HCOP{} shell
(Fig. \ref{fig:hcop}d).
The emission seen across the source may arise from both
the inner envelope and the shell, 
with the emission peak northwest of the source.

\subsubsection{Kinematics}

As mentioned, the \HCOP{} emission around the source perpendicular to the
jet axis may arise from the inner envelope.
A cut across the source perpendicular to the jet axis shows that
the emission is seen across the source with the
redshifted and blueshifted peaks on either side of the source
(Fig \ref{fig:pvhcop}a),
similar to that seen in the \cCO{} and \bCO{} emission 
\citep{Lee2006} but closer to the source tracing the inner envelope.
The blueshifted emission has a peak at $\sim$
($-1$ \vkm{}, $-$\arcsa{0}{05}) in the west, while 
the redshifted emission has a peak at
$\sim$ (1 \vkm, \arcsa{0}{05}) in the east.
This PV structure suggests that the inner
envelope is not only rotating around the
source but also infalling toward the source with the redshifted emission
from the nearside and the blueshifted emission from the farside
\citep{Lee2006}.
The emission is optically thick,
with the absence of the emission around the systemic velocity.  
The blueshifted emission is much brighter than the redshifted emission, 
consistent with a presence of an infall motion \citep{Evans1999}.
Note that the redshifted emission in the west at $\sim$ (0.7
\vkm{}, $-$\arcs{2}) may arise from the shell. 
On the other hand, the origin of the \aHCOP{} emission is 
unclear and may arise from both the inner envelope and the shell.

The kinematics of the shells can be studied with
the cuts across the shells centered at the jet axis.
A cut across the shell at \arcsa{2}{5} north of the
source shows two opposite \HCOP{} crescents at $\sim\pm$ 1 \vkm{} and one
\aHCOP{} structure around the systemic velocity (Fig. \ref{fig:pvhcop}b).
The two crescents may form the two ends of an elliptical structure connected
by the \aHCOP{} structure, suggesting that the shell there is expanding at
$\sim$ 1 \vkm{}. 
The \HCOP{} shell is probably optically thick, with the absence of the
emission around the systemic velocity.
The redshifted emission is much
brighter than the blueshifted emission, consistent with a presence of
an outflow motion.
Cuts across the shell at
\arcsa{1}{25} north (Fig. \ref{fig:pvhcop}c) and \arcsa{1}{25} south (Fig.
\ref{fig:pvhcop}d) of the source show a hint of rotation with
the redshifted emission to the
east and the blueshifted emission to the west, similar to that seen in the
inner envelope, suggesting that the shell is probably the material
extended from the inner envelope.
Note that, however, the center of
symmetry is shifted by $\sim$ \arcsa{0}{5} to the east in
the north, while \arcsa{0}{5} to the west in the south from the jet axis.

\subsubsection{Temperature, Column density, and Density}

In the following, we derive the excitation temperature, column density, and
density, assuming optically thin emission and local thermal equilibrium. Note
that due to the absorption of the ambient cloud, that part of the emission
is optically thick, and that part of the emission is resolved out by the
interferometer, the values of the column density and density
presented here are lower limits of the true values.

\begin{figure}[!hbp]
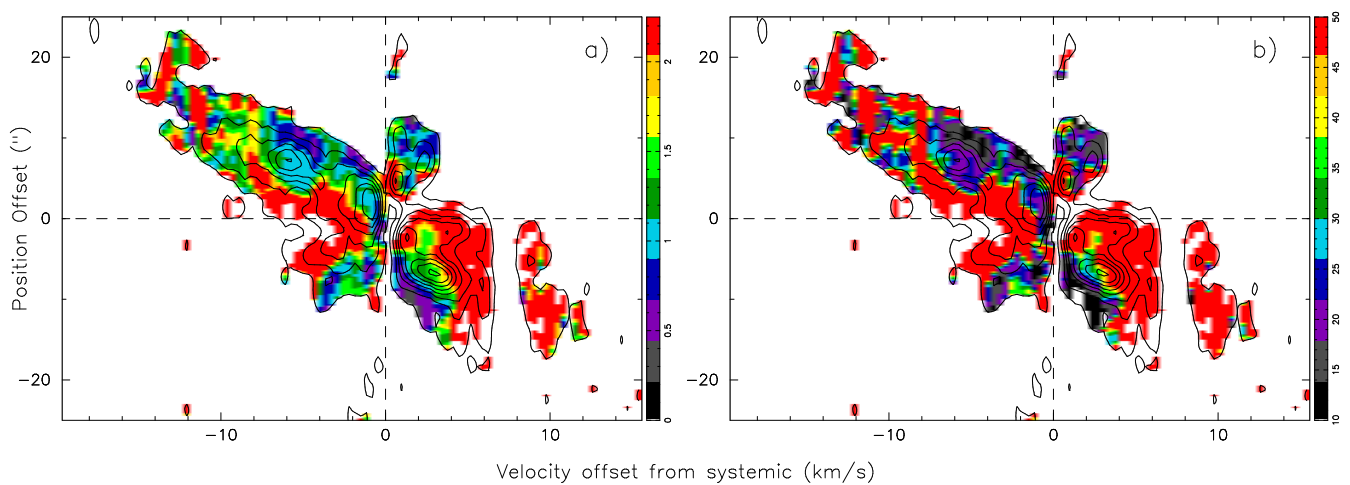

\centering
\putfig{0.7}{270}{f9.ps}
\figcaption[]{
\tlabel{a}
Line ratio of CO $J=3-2/J=2-1$ smoothed to the angular resolution in
the CO $J=2-1$ observations, which is $\sim$ \arcsa{2}{5}.
\tlabel{b} Excitation temperature derived from the line ratio.
The black contours in these two panels show the PV diagram of the 
CO $J=3-2$ emission cut along the jet axis, smoothed to the resolution 
of the CO $J=2-1$ observations.
\label{fig:COratio}
}
\end{figure}

The excitation temperature of the CO emission can be derived from
the line ratio of CO $J=3-2/J=2-1$ along the jet axis (Fig.
\ref{fig:COratio}a), using the CO $J=2-1$ observations from \citet{Lee2006}.
Note that, however, since the emission around 7 \vkm{} is resolved out more
in CO $J=2-1$, the line ratio there is not properly calculated and appears to
be larger than the maximum value of 2.25.
In addition, since some of the emission, mostly at high velocity, is too weak to
be detected in CO $J=2-1$, the line ratio there also appears to be larger than
the maximum value of 2.25.
At a given projected distance
from the source, the line ratio increases from low velocity to high
velocity, suggesting that CO $J=3-2$ traces high velocity (and thus jet)
better than CO $J=2-1$.
The temperature of the shells is found to be $\sim$ 20 K (Fig.
\ref{fig:COratio}b). The temperature of the high-velocity emission in the
jet is higher and assumed to be 50 K, with a peak brightness temperature
found to be $\sim$ 35 K in knot SS. 
The \H2{} column density can be derived assuming a CO
abundance of 8.5$\times10^{-5}$ \citep{Frerking1982}. In the jet, it is
found to be $\sim$ 0.4$\times 10^{21}$ \cms{} in the continuous
structures and $\sim$ 1.2$\times 10^{21}$ \cms{} in knots SS and SN.
Therefore, with
a diameter of $\sim$ \arcs{1}, the continuous structures has a density of
5.8$\times 10^{4}$ \cmc{}, and knots SS and SN have a density of
1.7$\times 10^{5}$ \cmc{}, similar to that found in the CO jet of HH
211 \citep{Gueth1999}.
On the other hand, the shells are found to have a column density of
$(2-7)\times 10^{20}$ \cms{} and thus
a density of $\sim$ $(0.3-1.0)\times10^5$ \cmc{},
with a shell thickness of $\sim$ \arcs{1}.

The excitation temperature of the SiO emission has been found to
be 50$-$150 K \citep{Gibb2004} and is thus assumed to be 100 K here.
This temperature, derived at much lower angular resolution,
can be considered as the lower limit of the true value.
With this temperature, the SiO column density is found to be
$1.5-2.5\times10^{14}$ \cms{} toward the bow shocks and knots.
The SiO abundance has been found to be 
$5\times 10^{-8}$ \citep{Gibb2004}. With a size $\lesssim$ \arcs{1}, 
the density is $\gtrsim(4-7)\times10^5$ \cmc{}.
Note that, since the temperature of the SiO emission
could be higher, i.e., 300-500 K as seen in HH 211 \citep{Hirano2006},
the density could be a factor of a few higher.

The excitation temperature of the SO emission can be 
derived from the line
ratio of SO \SOt{}/\SOta{}, using the SO \SOta{} observations from
\citet{Lee2006}. With a line ratio of $\sim$ 1.5, 
the excitation temperature is found to be $\sim$ 100 K,
similar to that found in the CepA-East outflows
\citep{Codella2005}.
The SO column density is found to be $\sim$ 8$\times10^{14}$ 
\cms{}. In shock model, the SO abundance is predicted to
be $3\times10^{-8}-6\times10^{-7}$ at 100 K \citep{Viti2002}.
Thus, with a size of $\sim$ \arcs{1}, the density is $2\times10^{5}-$
$4\times10^{6}$ \cmc{}.

The excitation temperature of the \HCOP{} emission
is close to the dust temperature, with a 
peak brightness temperature found to be $\sim$ 41 K toward the source.
Assuming a temperature of 50 K,
the \HCOP{} column density is found to be $\sim$ 3.2$\times10^{13}$ \cms{} 
toward the source and 1.8$\times10^{13}$ \cms{} toward the shells.
The abundance of \HCOP{} is uncertain. It is 
$\sim 2\times10^{-9}$ in molecular cloud cores \citep{Girart2000} but greatly
enhanced in outflow interacting regions by a factor of 20 with a value
of 4$\times10^{-8}$ \citep{Hogerheijde1998,Viti2002}. Therefore,
with a thickness of $\sim$ \arcs{1}, 
the density is $1.2\times10^5-2.4\times10^6$ \cmc{} toward the
source and $6.5\times10^{4}-1.3\times10^{6}$ \cmc{} toward the shells. 

The excitation temperature of the \aHCOP{} emission is unknown and assumed
to be the same as that of the \HCOP{} emission. The \aHCOP{} column density
is found to be $\sim$ $6\times10^{12}$ \cms{}. The \aHCOP{} abundance has
been found to be $\sim$ $10^{-10}$ in molecular cloud cores
\cite[see, e.g.,][]{Takakuwa2003}, but is unknown in
outflow interacting regions and could be enhanced by
a factor of 20, as the \HCOP{} abundance.
Thus, the density is $4\times10^{5}-8\times10^{6}$ \cmc{}, with a thickness
of $\sim$ \arcs{1}.

\section{Discussion}

\subsection{Molecular Jet}

\subsubsection{Shocks: SiO, CO, and \H2{}}

\begin{figure} [!hbp]
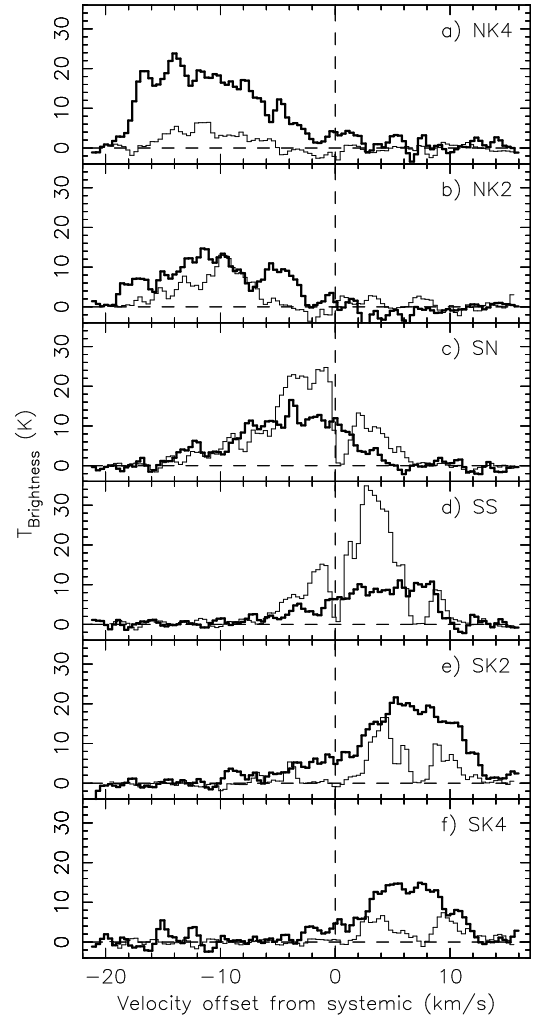

\centering
\putfig{0.8}{270}{f10.ps}
\figcaption[]
{SiO (thick lines) and CO (thin lines) spectra toward the bow shocks and knots.
\label{fig:specSiO}
}
\end{figure}

SiO is seen tracing the shocks in the jet,
as in HH 211 \citep{Palau2006}.
The emission is seen around the bow shocks with a range of velocities,
as predicted in the jet-driven bow shock models
\cite[see, e.g.,][]{Lee2001}.
Knots SN and SS are likely unresolved bow shocks, associated with a similar
range of velocities. 
\H2{} emission is also seen associated with them in the 
Spitzer IRAC 4 (8 \micron) and IRAC 3 (5.8 \micron) images.
No \H2{} emission is seen associated with
them in the VLT image at 2.12 \micron{} is
likely because of the severe dust extinction toward them at that wavelength.
The spectra toward bow shocks NK4, SK2 and SK4
and knot SS are asymmetric: steep toward the highest velocities with a wing
pointing toward the lowest velocities (Fig. \ref{fig:specSiO}), 
also as predicted in the shock model in which
the SiO abundance is enhanced more at higher shock velocity \citep{Schilke1997}.
It is believed that
SiO abundance is enhanced 
as a consequence of grain sputtering or grain-grain collisions releasing
Si-bearing material into the gas phase, which reacts rapidly with O-bearing
species (e.g., O$_2$ and OH) to form SiO
\citep{Schilke1997,Caselli1997}.
The emission is consistent with its production in C-type shocks, with a
total velocity range of $\sim$ 20 \vkm{} (see Fig. \ref{fig:specSiO}), 
similar to that found to produce
the observed SiO column densities in molecular outflows \citep{Schilke1997}.
The C-type shocks have also been argued to produce the high shock 
velocity gas seen in \H2{} \citep{Zinnecker1998}.
The SiO emission, with the derived
density orders of magnitude 
lower than the critical density, may arise
from ``thinner'' regions, as in thin shock fronts.

SiO may trace different shock conditions from \H2{}.
As mentioned, although \H2{} emission is bright around 
bow shocks NK1 and SK1, SiO emission is faint around them,
and is faintest around the former.
In \H2{}, the FWHM velocity widths toward the bow shocks
were found to be 9 (NK7), 6 (NK4), 12 (NK2), 31 (NK1), 21 (SK1), and 6 (SK2)
\vkm{} \citep{Davis2000,Takami2006}, indicating that
the shock is much stronger for bow shocks NK1 and SK1,
and is strongest for the former.
Thus, it is possible that SiO around bow shocks NK1 and SK1 
is mostly destroyed by the shocks through reactions such as
SiO$+$OH$\rightarrow$SiO$_2$$+$H \cite[see, e.g.,][]{Schilke1997}.
On the other hand,
SiO and \H2{} are known to trace different physical conditions,
with SiO ($T_\textrm{ex}\sim 100$ K, $n_\textrm{cr}\sim 10^8$ \cmc{})
tracing much cooler but much denser gas than \H2{} 
($T_\textrm{ex} \sim 2000$ K, $n_\textrm{cr}\sim 10^6$ \cmc{} for
\H2{}-\H2{} collision).
Thus, it is also possible that although gas-phase SiO is being produced,
the density there
is not high enough to produce bright collisionally excited SiO emission.
CO emission ($T_\textrm{ex} \sim 50$ K, $n_\textrm{cr}\sim 4\times10^4$ \cmc{}),
which traces cooler and less dense gas than SiO emission, 
is seen around there, also supporting this possibility.
This possibility may apply as well to weak bow shocks such as NK6, NK7, 
and SK5.
For those weak bow shocks, however, it is also possible that SiO there
is depleted back onto grains in the postshock gas.

The density in the bow shocks 
is expected to decrease with the distance from the source
due to the sideways ejection of the shocked material and
the possible thermal expansion of the jet. The fact that the CO emission
decreases with the distance from the source 
(see Fig. \ref{fig:specSiO}) also supports this possibility.
However, due to the shock enhancement, the SiO emission is brightest around
bow shocks NK4 and SK2. 

CO emission is also seen tracing the shocks in the jet.
For those bow shocks, e.g., NK4 and SK4, where CO and SiO emission are
coincident, the velocity dispersion is smaller in CO than SiO, 
suggesting that CO traces weaker shock than SiO.
It is likely because CO traces cooler and less dense gas than SiO, and thus
further away from the bow tip where the shock velocity is lower.

\subsubsection{Inclination}

The jet inclination, $i$, can be estimated from the mean velocity of the SiO
emission, $v_m$, with $i = \sin ^{-1} (v_m/v_j)$, where $v_j$ is the jet
velocity. 
The northern component and southern component of the jet, however,
may have different inclinations because of their different mean velocities.
Assuming the same jet velocity,
the northern component and
southern component of the jet have an inclination of
$\sim$ 6.3\degree{}$(100\; \vkme/v_j)$ and
4.0\degree{}$(100\; \vkme/v_j)$, respectively, in agreement with that found
from the water masers \citep{Claussen1998}. Note that, knots SS and SN may
have lower inclinations, because of their lower mean velocities.

\subsubsection{Mass-loss rate}

Continuous structures between the bow shocks may trace the
``intrinsic'' jet itself, 
allowing us to estimate the mass-loss rate of the jet,
which is given by
\begin{equation}
\dot{M}_j \sim n_c \frac{\pi d_c^2}{4} v_j m_{\textrm{\scriptsize H}_2}
\end{equation}
where $n_c$ and $d_c$ are the number density and the diameter of
the continuous structures.
With $n_c \sim 5.8\times 10^{4}$ \cmc{}, $d_c \sim 460$ AU, (i.e., 1\arcsec{}), and
$v_j=100-200$ \vkm{}, the mass-loss rate is
$(1-2)\times 10^{-6}$ \solarmass{} yr$^{-1}$, about 15\%$-$30\%
of the infall rate derived from the envelope \citep{Lee2006}.

\subsubsection{Episodic nature}

In the inner part of the jet, a chain of bow shocks are seen with
a semiperiodic spacing of $\sim$ \arcs{3}.
Between the bow shocks, the continuous structures are
seen probably with the velocity increasing periodically
from one bow shock to the next,
suggesting that the jet velocity varies periodically with time and is
highest at the far ends of the continuous structures.
These are consistent with a pulsed (i.e., episodic) 
jet model with a periodical velocity variation, in which
a chain of bow shocks are formed as the fast-moving jet material
impacts on the slow-moving jet material \cite[see, e.g.,][]{Suttner1997,Lee2001}.
The jetlike SO emission around the source,
where the velocity probably has not steeped into shock,
may show intrinsically how the velocity varies with time.
Observations at higher angular resolution are needed to resolve it.

The period of the velocity variation, which can be estimated by dividing 
the semiperiodic spacing between the bow shocks by the jet velocity,
is found to be $P\sim 65(100\; \textrm{\vkm}/v_j)$ yr.
The jet is believed to be launched from an accretion disk around the source.
The periodic velocity variation may be
due to periodic perturbation of the accretion disk in an eccentric binary system;
in this case the period of the velocity variation is equal to the
orbital period of the binary and the
separation of the binary is given by
\begin{equation}
a=\big(\frac{GMP^2}{4\pi^2}\big)^{1/3}
\end{equation}
where $M$ is the total mass of the binary.
With $M\sim$ 0.15 \solarmass{} \citep{Lee2006} and $P=65$ yr, 
we have $a \sim$ 8 AU (i.e., \arcsa{0}{02}), similar
to that estimated in HH 34 \citep{Reipurth2002}. A binary with this
separation, however, would be unresolved in our observations.
The periodic variation may also be due to FU Ori
like events as the disk builds up mass from the envelope then dumps
it onto the protostar in a burst.

\subsubsection{Precessing}

The jet itself may be slightly precessing as the jetlike SO emission
shows a slight S-shaped symmetry about the source.
The fact that the continuous structures seen in \H2{} and CO
are sinuous also supports 
this possibility. 
The angle of precession can be assumed to be the
angle subtended by the continuous structures 
and is found to be $\lesssim$ 2\degree{}.
Since the continuous structures may show a full cycle between 
bow shocks SK1 and SK2, and between bow shocks NK1 and NK2, the
period of the precession is probably similar to that of the
velocity variation, suggesting that the precession is also 
due to a binary system.
A slight S-shaped symmetry is also seen in other jets, e.g., HH 34,
and is generally ascribed to the tidal
effects of the companion star on the direction of the jet axis
\citep{Reipurth2002}. However,
it is also possible that the slight precession is due
to kink instability in the jet \citep{Todo1993}

\subsubsection{Bending}\label{sec:dis_bend}
The inner part of the jet is bent by $\sim$ 1\degree{} to the west. Similar
jet bending is also seen in HH 211 \citep{Gueth1999}. 
It has been proposed that a jet can be bent (or deflected)
by the dynamical pressure of the ambient medium \citep{Fendt1998}.
However, it is not clear how the ambient medium
can be communicated to the highly supersonic jet.
%The lack of \H2{} emission in the northwest of the source could be
%because the source has cleared up the material there as it moved.

The bending could be due to motion of the jet source in a binary
system \citep{Fendt1998}.  In this case, a binary system with
a minimum separation of $\sim$ \arcsa{0}{5} (or 230 AU), which is the lower limit set by
the half displacement of bow shocks NB1/2 and SB1/2 from the original axis,
is required.
However, no binary companion has been detected yet. It is possible that
the bending is due to Lorentz force on the magnetic jet
\citep{Fendt1998}.
It is also possible that the jet can appear to be bent just due to the
motion of the star/disk system, i.e. the jet is originating from different
positions as a function of time due to the star's peculiar velocity.

\subsubsection{Rotation?}

\begin{figure}[!hbp]
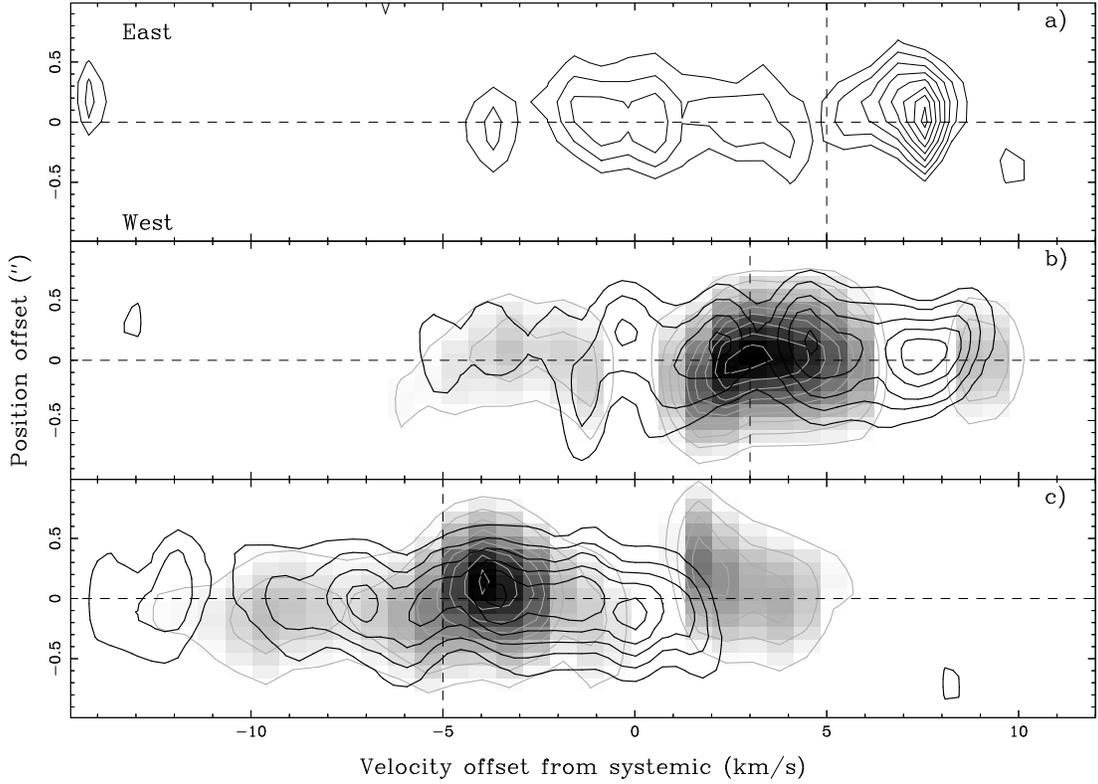

\centering
\putfig{0.6}{270}{f11.ps}
\figcaption[]{
PV diagrams with cuts across 
\tlabel{a} the southern component of the SO jetlike structure,
\tlabel{b} knot SS, and \tlabel{c} knot SN.
The cuts have a width of \arcsa{0}{15}.
The dashed vertical lines indicate the possible jet central velocities 
for the cuts.
In \tlabel{b} and \tlabel{c}, the
black contours are from the SiO emission and the gray-scale image with gray
contours is from the CO emission.
\label{fig:pv-rotate}
}
\end{figure}

In theoretical jet-launching models, the jet is expected to be rotating,
carrying away angular momentum from the infalling envelope and the accretion
disk.
The far end of the southern component of the SO jetlike structure,
which is at a beam size ($\sim$ \arcs{1}) away from the source,
is likely far enough from the
possible disk contamination around the source and thus can be used to study
the jet rotation.
However, no clear velocity gradient is seen across it
(Fig. \ref{fig:pv-rotate}a).
Knots SS and SN are the shocks closest to the source and
can also be used to study the jet rotation.
Although no clear velocity gradient is seen across knot SS (Fig.
\ref{fig:pv-rotate}b),
a velocity gradient is seen across knot SN 
(Fig. \ref{fig:pv-rotate}c) with the redshifted
emission to the east and the blueshifted emission to the west, similar to
that seen in the envelope.
Thus, jet rotation may be seen in our observations.
Observations at higher angular resolution are needed to
confirm this.

\subsection{Molecular Outflow Shells} \label{sec:shells}

Outflow shells are seen in CO $J=3-2$ surrounding the jet extending to the
north and south from the diffuse nebulae 
that trace the outflow cavity walls (MZAML02),
similar to that seen in CO $J=2-1$ \citep{Lee2006}.
However, the shells are asymmetric in CO $J=3-2$, with the northern shell
bright in the west and southern shell bright in the east. Since CO $J=3-2$
traces warmer material and thus more recent outflow interactions than CO
$J=2-1$, this asymmetry indicates that the recent outflow interactions are
asymmetric. This can be explained if the jet axis has rotated in the
clockwise direction recently, 
as suspected from the morphology of the \H2{} jet (see \S
\ref{sec:jetaxis}). 

Around the source, internal outflow
shells are seen in CO and \HCOP{} associated with, 
e.g., bow shocks SK1 and NK1.
They probably trace the material squirted out from the bow
shocks, as in pulsed jet model \cite[see, e.g.,][]{Suttner1997,Lee2001}.
The shells are asymmetric, with the
northern shell bright in the east and the southern shell bright in the west.
This is expected if the shells are produced by the slight S-shaped jetlike
SO structure. 
The bases of the \HCOP{} shells, which are seen
with a hint of rotation, however, are probably consisted mainly of the 
material extended from the inner envelope and
even the possible disk.
The bases are also seen in \aHCOP{} and continuum,
probably tracing the dense material extended from around the same
regions.

These internal outflow bases 
may trace the material entrained from the
inner envelope by the jet. However, in the
currently existing jet simulations
\cite[see, e.g.,][]{Suttner1997,Downes1999,Lee2001,Raga2004},
no internal outflow bases have been seen
because the jet is protected by a slower sheath shock 
from continued interaction with the ambient material
at its bases. It is possible that these internal outflow
bases trace the material
directly launched from the inner envelope (and even the possible disk)
around the jet and then swept up by the internal bow shocks, e.g., NK1 and SK1, 
of the jet. 
Further simulations are needed to study this.

\subsection{Wide-angle wind component?}
The jet itself is clearly highly collimated, yet the diffuse
nebulae clearly show the classic bipolar parabolic shape with a wide opening
angle. This may suggest that an unseen wide-angle wind component is there to
produce the diffuse nebulae, so that the jet is only a dense component of a
wider wind. However, as seen in Figure \ref{fig:jet}a, the jet produces a
chain of bow shocks with the interacting surface growing with the distance
from the source, and thus producing a few prominent bow shocks 
with much bigger interacting surfaces than the jet itself. Through these bow
shocks, a jet may be able to produce wide-opening outflow bases such as the
diffuse nebulae. Detailed modelings are needed to check this.

\subsection{Molecular Envelope and Disk?}

\begin{figure}[!hbp]
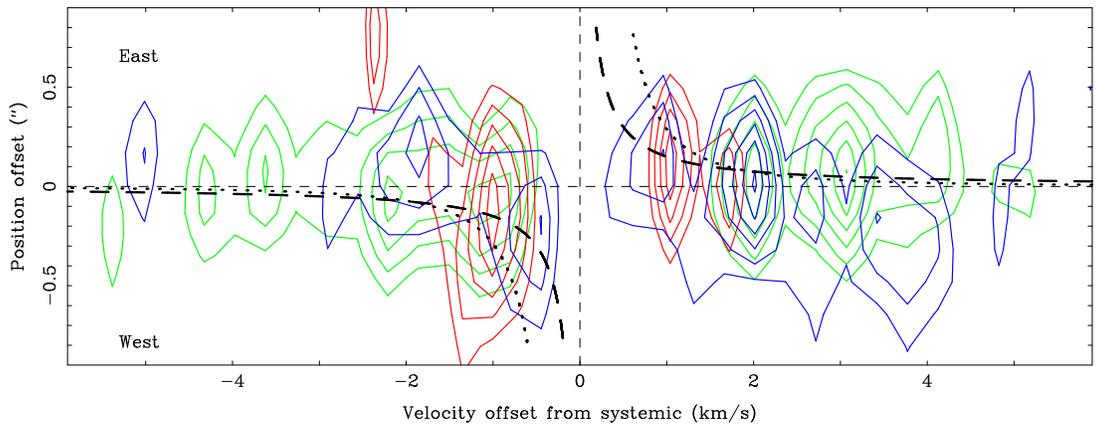

\centering
\putfig{0.6}{270}{f12.ps}
\figcaption[]{
PV diagrams of CO (green), \HCOP{} (red), and SO (blue) emission
cut across the source. The dotted curves are calculated assuming a Keplerian
rotation with a source mass of 0.15 \solarmass{}. The dashed curves are
calculated assuming a constant specific angular momentum
with $v_\textrm{rot} = \frac{0.15}{d}$ \vkm{}, with $d$ being the
projected distance in arcsec. 
Note that the fits could be affected by the missing flux around the systemic
velocity.
\label{fig:pvdisk}
}
\end{figure}

The \HCOP{} emission around the source may arise from the inner part of a
dynamically infalling envelope with rotation,
showing a kinematics similar to that seen in the \cCO{} and \bCO{} emission
in \citet{Lee2006}, but closer to the source.
As argued in \citet{Lee2006}, a compact rotationally supported disk with a
radius of $\sim$ 74 AU (or \arcsa{0}{16}) is
expected to be formed within the inner part of the envelope.
In order to study it, the PV diagrams of \HCOP{}, CO, and SO emission
cut across the source are compared
with a Keplerian rotation law 
\cite[with a source mass of
0.15 \solarmass{},][]{Lee2006} 
and a rotation law with a constant specific angular momentum
(Fig. \ref{fig:pvdisk}).
In the diagrams,
the large-scale emission structures are excluded (see \S \ref{sec:obs}), in
order to study the inner regions.
Note that the comparison could be affected by the missing flux around the systemic
velocity.
Rotation is clearly seen within 100 AU from the source in CO and \HCOP{}, 
with the redshifted
emission on the east and the blueshifted emission on the west. 
The SO emission may have two components, with the low velocity
($\lesssim 2$ \vkm{} from the systemic velocity) from the rotation and the
high velocity from the jet. The PV structures seem better fitted by the
rotation law with a constant specific angular momentum, indicating that
the emission within 100
AU from the source probably is still from the inner part of a dynamically
infalling envelope with rotation rather than a rotationally supported disk.
The rotation may change to Keplerian in the innermost region.
Observations at higher angular resolution and detailed modeling are
both needed to check this.
%Note that we only see rotation in small scale in \HCOP{}, 
%suggesting that the infall motion appears mainly in larger scale.

\section{Conclusions}
We have mapped the protostellar jet HH 212 in 850 \micron{} continuum, SiO
$J=8-7$, CO $J=3-2$, SO \SOt{}, \HCOP{}$J = 4-3$, and \aHCOP{} $J = 4-3$
emission.
Thermal dust emission is seen in continuum around the source IRAS 05413-0104, 
mainly arising from an inner envelope
and a possible disk.
The inner envelope is also seen with rotation in CO, 
\HCOP{}, and probably SO.
The structure is unresolved but likely to be flattened
perpendicular to the jet axis, as seen in the CLEAN component map.
Like \H2{} emission, CO and SiO emission are seen
along the jet axis but extending closer to the source,
tracing the bow shocks with a broad range of velocities and
the continuous structures in between.
SO emission is seen only around
the source, forming a jetlike structure extending along the jet axis 
from the source, likely tracing the jet near the launching region.
The jet is episodic and bending. It may also be slightly precessing as
the jetlike SO structure shows a slight
S-shaped symmetry about the source. A hint of jet rotation is also
seen across the jet axis. 
Internal outflow shells are seen in CO and
\HCOP{}, associated with the bow shocks in the inner part of the
jet. The bases of the \HCOP{} shells are seen with a hint of rotation
similar to that seen in the inner envelope, probably consisted mainly of the
material extended from the inner envelope and even the possible disk.
The bases of the outflow shells are also seen
in \aHCOP{} and even the continuum, probably tracing the dense
material extended from around the same regions.
Outflow shells are also seen in CO surrounding the jet extending
out from the \H2{} nebulae. They are probably internal outflow shells 
associated with the bow shocks further down the jet axis.

\acknowledgements
We thank the anonymous referee for insightful comments.
We thank the SMA staff for their efforts
in running and maintaining the array.
H.B. acknowledges financial support by the Emmy-Noether-Program of the
Deutsche Forschungsgemeinschaft (DFG, grant BE2578).

%% Remember to include "(" and ")" for the year,e.g., (1998)
%%

%critical density is given by Ajk/\gammajk
%
\begin{deluxetable}{lcccc}
\tablecolumns{6}
\tabletypesize{\normalsize}
\tablecaption{Line properties
 \label{tab:obs}}
\tablewidth{0pt}
\tablehead{
\colhead{Line} & \colhead{Frequency} & $E_\textrm{up}$&  $n_{cr}$(50-100 K)  \\
               & \colhead{(GHz)}     & (K)            &   (\cmc{}) }  
\startdata
%850 \micron{} continuum & 351.500000  &    &  \\
SiO $J=8-7$       & 347.330631 & 75.0 & 1.1$\times10^8$ \\
SO \SOt{}         & 346.528481 & 78.8 & 1.4$\times10^7$ \\
CO $J=3-2$        & 345.795991 & 33.2 & 3.6$\times10^4$ \\
\HCOP{} $J=4-3$   & 356.734288 & 42.8 & 9.0$\times10^6$\\
\aHCOP{}$J=4-3$   & 346.998338 & 41.6 & 8.1$\times10^6$\\
\enddata
\tablenotetext{\mbox{}}{$n_{cr}$: Critical density of molecular hydrogen.
%using Leiden Atomic and Molecular Database.
}
\end{deluxetable}
\clearpage

\begin{figure} [!hbp]
\centering
%\putfig{0.85}{270}{f2.ps}
\figcaption[]
{ 
\tlabel{a} The \H2{} image adopted from MZAML02.
The ellipse outlines the observed region in our observations.
The dotted lines indicate the jet axes.
The cross marks the source position.
\tlabel{b} The \H2{} image rotated by 22.5\degree{} clockwise.
\tlabel{c} SiO (integrated from -21.1 to 16.0 \vkm{})
contours plotted on top of the \H2{} image.
The contours go from 3 to 21 $\sigma$ with a step of 2 $\sigma$, 
where $\sigma = 0.58$ \Jybk{}.
\tlabel{d} High-velocity CO (integrated from -18.4 to -7.1 and from 3.4 to
13.3 \vkm{}) contours plotted on top of the \H2{} image.
The contours go from 3 to 21 $\sigma$ with a step of 3 $\sigma$, 
where $\sigma = 0.38$ \Jybk{}. 
The synthesized beams are 
\arcsa{0}{96}$\times$\arcsa{0}{69} and 
\arcsa{1}{16}$\times$\arcsa{0}{84}, respectively, for the SiO and CO emission.
\label{fig:jet}}
\end{figure}

\begin{figure} [!hbp]
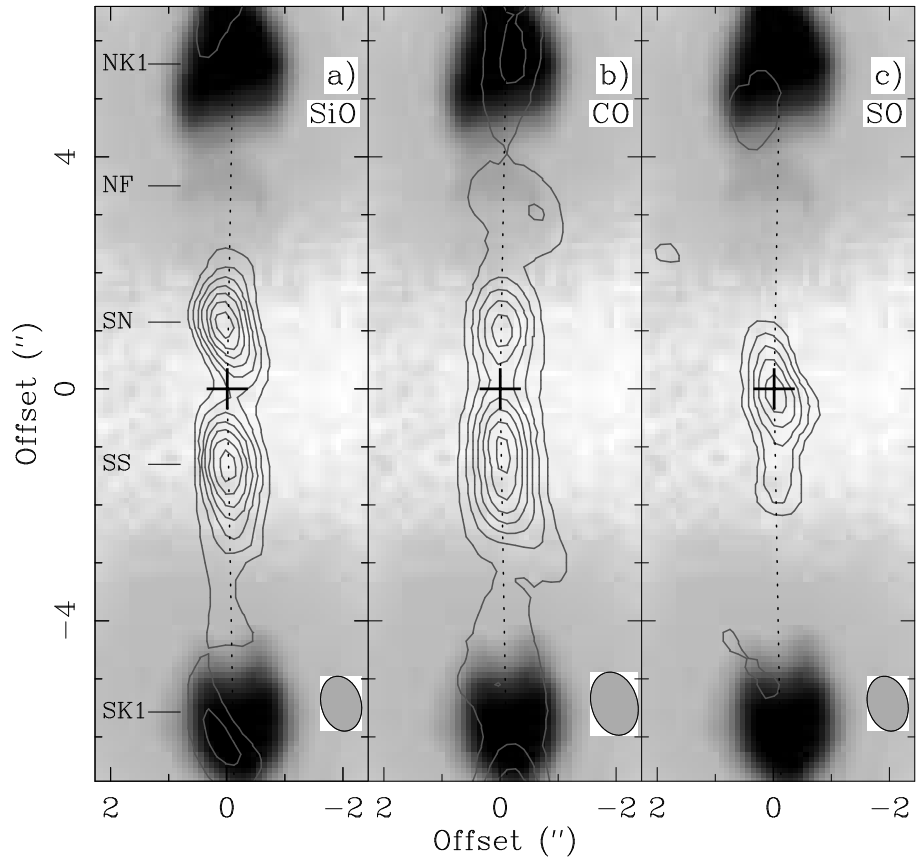

\centering
\putfig{0.8}{270}{f3.ps}
\figcaption[]
{\tlabel{a} SiO, \tlabel{b} High-velocity CO, and \tlabel{c} SO contours
on top of the \H2{} image near the source. 
The SiO and CO contour levels are
the same as those in Fig. \ref{fig:jet}c and \ref{fig:jet}d, respectively.
The SO (integrated from -11.8 to 11.1 \vkm{})
contours go from 2 to 10 $\sigma$ with a step of 2 $\sigma$,
where $\sigma = 0.45$ \Jybk{}.
The synthesized beams are 
\arcsa{0}{96}$\times$\arcsa{0}{69} for the SiO and SO emission and 
\arcsa{1}{16}$\times$\arcsa{0}{84} for the CO emission.
\label{fig:injet}
}
\end{figure}

\begin{figure}[!hbp]
\centering
\putfig{0.8}{270}{f7.ps}
\figcaption[]{
\HCOP{}, continuum, and \aHCOP{} contours plotted on top of the \H2{} image.
The synthesized beams all have a size
of \arcsa{1}{16}$\times$\arcsa{0}{84}.
\tlabel{a} \HCOP{} (black, integrated from -2.5 to 2.5 \vkm{}) 
and low-velocity CO (green)
contours. The \HCOP{} contours go from 3 to 15 $\sigma$ with a step of 
2 $\sigma$,  where $\sigma = 0.32$ \Jybk{}.
\tlabel{b} Blueshifted (integrated from -2.5 to 0 \vkm{})
and redshifted (integrated from 0 to 2.5 \vkm{}) \HCOP{} emission, with
the contours from 3 to 11 and 3 to 9 $\sigma$, respectively, 
with a step of 2 $\sigma$,  where $\sigma = 0.23$ \Jybk{}.
\tlabel{c} \HCOP{} (black) and continuum (orange) 
contours.
\tlabel{d} \HCOP{} (black) and \aHCOP{} (cyan, integrated from -1.7 to 2.5
\vkm{}) contours.
The \aHCOP{} contours go from 3 to 6 $\sigma$ with a step of 
1 $\sigma$,  where $\sigma = 0.20$ \Jybk{}.
The horizontal dashed lines indicate the cuts for the PV diagrams 
in Fig. \ref{fig:pvhcop}.
\label{fig:hcop}
}
\end{figure}

\begin{figure}[!hbp]
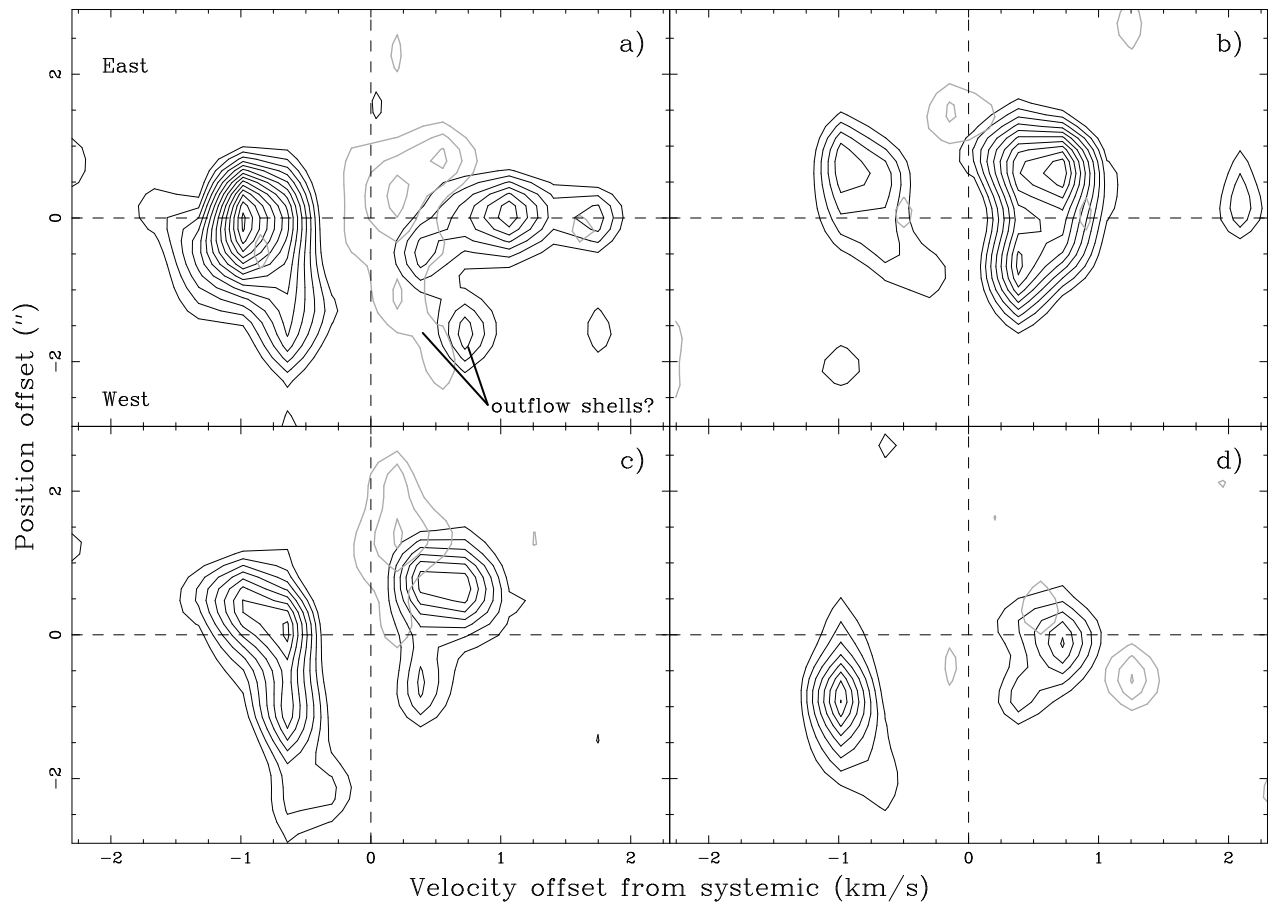

\centering
\putfig{0.7}{270}{f8.ps}
\figcaption[]{
PV diagrams of the \HCOP{} (black) and \aHCOP{} (gray)
emission cut across the jet axis, with
the cuts shown in Fig \ref{fig:hcop}d.
\tlabel{a} A cut centered at the source.
Cuts centered at 
\tlabel{b} \arcsa{2}{5} to the north,
\tlabel{c} \arcsa{1}{25} to the north, and
\tlabel{d} \arcsa{1}{25} to the south, respectively, of the source.
\label{fig:pvhcop}
}
\end{figure}

\end{document}